**What controls the superconducting dome of electron-doped FeSe?**

Paul T. Malinowski[1], Chad J. Mowers[1], Yaoju Tarn[1], Darrell G. Schlom[2,4,5], Brendan D. Faeth[1,3], and Kyle M. Shen[1,4,*]

[1]Department of Physics, Laboratory of Atomic and Solid State Physics, Cornell University, Ithaca, NY, 14853, USA

[2]Department of Materials Science and Engineering, Cornell University, Ithaca, New York, 14853, USA

[3]Platform for the Accelerated Realization, Analysis, and Discovery of Interface Materials (PARADIM), Cornell University, Ithaca, NY, 14853, USA

[4]Kavli Institute at Cornell for Nanoscale Science, Cornell University, Ithaca, NY, 14853, USA

[5]Leibniz-Institut für Kristallzüchtung, Max-Born-Str. 2, 12489 Berlin, Germany

***Corresponding author**: Kyle M. Shen
**Email:** kmshen@cornell.edu

**Abstract**

Superconducting domes are conspicuous features of the phase diagrams of most unconventional and high-temperature superconductors. The superconducting transition temperature ($T_c$) of FeSe can be dramatically enhanced with electron doping, but unlike all other high-temperature and unconventional superconductors, its full phase diagram and superconducting dome have yet to be fully explored. Here, we employ a combination of molecular beam epitaxy synthesis, alkali surface doping, in-vacuum electrical transport, and angle-resolved photoemission spectroscopy to investigate the entire superconducting dome of electron-doped FeSe, achieving a fully metallic state where superconductivity is suppressed in the heavily overdoped regime. We discover a robust scaling between $T_c$ and the residual resistivity ($\rho_0$) which holds across the entire superconducting dome, suggesting that the evolution of $T_c$ is heavily influenced by the evolution of the elastic scattering rate in the high-$T_c$ electron-doped phase. This in turn suggests that the superconducting dome in electron-doped FeSe appears to be fundamentally different than that of other unconventional superconductors where doping plays the primary role, and may be driven primarily by the sensitivity of the superconductivity to disorder.

**Introduction**

Among unconventional superconductors, a common hallmark is that the superconducting transition temperature ($T_c$) exhibits a dome as a function of some tuning parameter such as doping or pressure. The careful investigation of these domes has provided crucial insights into the origin of superconductivity and its relation to properties of the normal state. For instance, in Fe-pnictide[1], cuprate[2], and heavy fermion[3] superconductors, the optimal $T_c$ often appears proximate to an enigmatic strange metal phase[4] with linear-in-temperature electrical resistivity and a critical fan emanating from a putative quantum critical point, whereas Fermi liquid behavior is recovered in the overdoped regime. These interdependencies between superconductivity and the normal state provide critical information about the pairing mechanism and the nature of superconductivity. For FeSe, $T_c$ is enhanced with electron doping, but unlike all other high-$T_c$ and unconventional superconductors, its full phase diagram and superconducting dome have yet to be fully explored due to the difficulty in continuously and uniformly doping the material up to high carrier concentrations past optimal doping.

Bulk, undoped FeSe has a relatively modest $T_c \sim 8$K, however when doped with electrons so that the hole pockets at the Brillouin zone center ($\Gamma$) are pushed below the Fermi level ($E_f$), $T_c$ can be substantially enhanced. This doping can be achieved in a variety of ways, such as liquid[5–9] or solid[10–12] ionic gating, electrochemical etching[13–16], and the intercalation of alkali metals[17–22] or organic materials[23,24] into single crystals, where values of $T_c$ near 40 K have been reported. In monolayer FeSe / $SrTiO_3$, doping is realized via interfacial charge transfer, resulting in a zero resistance state approaching $T_c \sim 30$ K[25] and a spectroscopic gap opening below $T_\Delta \sim 70$ K[25–28].

Despite this substantial body of work establishing electron doping as a key ingredient for high $T_c$'s in FeSe, there have not yet been systematic studies spanning the entire superconducting dome in a continuous fashion, for several reasons. First, in protonated or intercalated systems such as $A_xFe_{2-y}Se_2$ ($A$ = alkali metal), the doping is difficult to control precisely and cannot be varied in a continuous or uniform fashion. Instead, only certain, discrete doping concentrations are stable, which typically form different crystallographic structures, and at other concentrations the dopants microscopically phase separate into insulating, semiconducting, or discretely disconnected superconducting regions[9,11,12,20,29,30]. On the other hand, electrostatic gating experiments can eventually cause irreversible damage to the sample at high voltages / doping levels[5,7], leaving the overdoped region of the phase diagram inaccessible. Surface alkali doping can resolve these issues, allowing for uniform, precise, and continuous doping of the sample up to very high levels, but must be performed in ultrahigh vacuum on atomically clean surfaces. However, to date, no *in situ* electrical transport measurements have been performed on surface-doped materials. As the spectroscopic gap measured by ARPES or tunneling measurements only reflects local pairing which is known to occur at higher temperatures than the thermodynamic superconducting transition in this material[25], the use of electrical transport is essential for tracking the true, phase-coherent superconducting $T_c$ across the phase diagram.

Here, we comprehensively probe the entire superconducting dome of electron-doped FeSe, spanning from the undoped parent compound all the way into the heavily overdoped, non-superconducting, metallic regime, combining surface electron doping of thin films of FeSe with *in situ* electrical transport and ARPES measurements. We discover an unusual non-monotonic dependence of the extrapolated residual resistivity, $\rho_0$, which reaches a minimum at precisely optimal doping, and find that $T_c$ scales precisely with $\rho_0$ across the entire dome on both the under- and overdoped sides. This unique behavior, where $\rho_0$ determines $T_c$ in spite of the changing carrier density, electronic structure, and other variables tuned by doping, suggests that disorder, not carrier doping, plays a dominant role in controlling the superconducting dome of electron-doped FeSe, in contrast to other unconventional superconductors such as the cuprates, Fe pnictides, and infinite-layer nickelates. While electron doping is necessary to initially reach the high-$T_c$ phase, our results suggest that once this minimum doping level is achieved, further changes in carrier density play a secondary role. This result places constraints on the microscopic aspects of superconductivity in this system, including shedding light on the as yet undetermined symmetry of the superconducting state of electron-doped FeSe, providing evidence for a sign-changing $s^{\pm}$ superconducting order parameter.

**Results**

In Fig. 1, we show resistivity measurements taken *in situ* on a 10-unit cell (u.c.) thick FeSe thin film as a function of Cs deposition. A key for this experiment was the precise control over the doping, which was achieved by evaporating Cs from a novel Cs-In alloy source[31] in an effusion cell. Due to the volatility of the adsorbed Cs, measurements were performed below 80 K, where the resistivity is repeatable upon temperature cycling and no Cs migration or loss occurs (Methods and SI Appendix, Fig. S4). Although bulk FeSe superconducts below $T \sim 8$ K, thin films on $SrTiO_3$ show little to no signature of the bulk superconducting phase, possibly due to epitaxial strain from the substrate[32]. Such films can exhibit superconductivity due to the singular interfacial FeSe layer, but here we deliberately omit the postgrowth annealing step essential for developing interfacial high-$T_c$ superconductivity (SI Appendix, Figs. S1-S2), allowing us to probe only the superconductivity that develops due to the surface doping of the topmost layer.

In Fig. 1b, we show data spanning from undoped to optimal doping, and in Fig. 1c spanning from optimal doping to overdoped. Upon depositing $x = 0.03$ Cs atoms per surface Fe atom, as determined by quartz crystal microbalance (QCM) measurements, a clear superconducting transition develops around $T_c = 15$ K, accompanied by a drop in the normal state resistance ($R_N$). Increasing $x$ further raises $T_c$ and continues to lower $R_N$, until the maximum $T_{c,opt} = 36.5$ K is reached at an optimal doping level of $x_{opt} = 0.076$. When increasing $x$ beyond optimal doping (Fig. 1c), $T_c$ is smoothly suppressed and eventually extinguished around $x \sim 0.24$. Upon crossing $x_{opt}$, $R_N$ begins to increase monotonically with $x$ on the overdoped side of the phase diagram. To better visualize the doping / temperature phase diagram, we plot the resistivity as a color map in Fig. 1d and define two characteristic temperatures, $T_{c0}$ (zero resistivity) and $T_{ons}$ (transition onset) (SI Appendix, Fig. S5), which are also plotted in Fig. 1d. The non-monotonic evolution of $R_N$ with $x$ is apparent as a region of low resistance (green color, white dashed lines) which exhibits a larger dome-like structure above the superconducting dome, indicating a clear correlation between the superconductivity and the normal state resistivity.

The data presented in Fig. 1 is the full sheet resistance of the entire 10 u.c. thin film, but the charge transfer of the deposited Cs is confined almost entirely to the topmost FeSe layer (SI Appendix). This has been clearly established by prior studies using surface alkali metal deposition on FeSe[33,34] and is also shown in our experiments. ARPES measurements as a function of doping (Fig. 2, SI Appendix, Fig. S3) highlight the changes in Fermiology induced by Cs deposition. As has been previously reported, a Lifshitz transition occurs at $x \sim 0.02$, before the onset of the superconducting dome, when the hole pockets at Γ are pushed below the Fermi level, leaving only electron pockets at *M.* With further doping, the electron pockets at *M* grow smoothly and monotonically, passing through optimal doping and well into the overdoped regime without any qualitative changes in the Fermiology.

Spectra at Γ (Fig. 2c-d; SI Appendix, Fig. S3) also show that the electron doping is confined to the first surface layer, as the bulk-like, undoped hole bands of the second layer do not shift in energy, regardless of the amount of Cs deposited, while the hole bands of the doped topmost layer continuously move to higher binding energy (Fig. 2f). These measurements also demonstrate that the change in Luttinger volume of this topmost layer, as measured by the size of the electron pockets at the *M* point (Fig. 2e,g), scales 1:1 with the amount of Cs deposited on the surface, determined independently by QCM measurements of the Cs flux. Furthermore, in Figs. 1 and 2, $x$ is plotted assuming the doped electrons are confined to the topmost FeSe layer, and these values of $x$ are consistent with previous studies[33–35].

Since only the doping of the topmost layer is changing with Cs deposition, we extract the sheet resistivity of the top layer alone by modeling the system as a stack of resistors with a doped top layer in parallel with 9 layers of undoped FeSe (see Methods). This is shown in Figs. 3a,b and is qualitatively like that of the entire film shown in Figs. 1b,c, but the doping-induced changes to the top layer are more dramatic. In Fig. 3c we plot the extrapolated residual resistivity ($\rho_0$) (SI Appendix, Fig. S6) of the doped layer which evolves non-monotonically, dropping over 80% from its undoped value $\rho_0 = 4.1$ kΩ to a minimum $\rho_0 = 760$ Ω at $x_{\mathrm{opt}}$, and then returning to $\rho_0$ = 4.2 kΩ at the highest doping of $x$ = 0.26. We note that this drop in the resistivity is unlikely to be attributable to superconducting fluctuations which manifest as small, temperature-dependent deviations from the normal state resistivity, and only become significant close to $T_{c0}$. The behavior we observe instead demonstrates a nearly vertical (i.e., temperature independent) offset of the curves for different doping levels, indicative of a change in the elastic scattering rate and reminiscent of Matthiessen's rule for point defects. Furthermore, the magnitude of the change (~ 80%) as well as the temperature range over which the drop is observed ($T > 2T_{c0}$) clearly suggests that conventional paraconductivity cannot be the origin of the doping dependence of $\rho_0$.

A comparison of the behavior of $\rho_0$ to the superconducting dome in Fig. 3d reveals that the minimum in $\rho_0$ occurs precisely at $x_{opt}$. Moreover, a comparison of underdoped and overdoped curves with similar values of $T_{c0}$ (Fig. 3e) show that each pair of curves are strikingly similar with nearly identical values of the normal state resistance, indicating a clear correlation between $T_c$ and $\rho_0$ on both sides of the superconducting dome. This is made evident in Fig. 3g, where plotting $T_{c0}$ as a function of $\rho_0$ reveals a direct linear scaling of these two quantities, with both underdoped and overdoped data points lying on a single line. We note this relationship also holds for the raw resistivity curves shown in Fig. 1 and also applies when comparing the normal state resistivity to $T_{c0}$, e.g. plotting $T_{c0}$ vs. $R_N$(50 K), and is largely insensitive to which temperature the resistivity is sampled (SI Appendix, Fig. S7). Furthermore, this behavior has been reproduced on multiple samples, demonstrating that this behavior is robust and reproducible (SI Appendix, Fig. S9). In addition, this relationship between $T_c$ and $\rho_0$ holds even when considering the raw resistivity for the full 10 u.c. thick FeSe film (SI Appendix, Fig. S8). These considerations demonstrate that this correlation between $T_c$ and $\rho_0$ is an intrinsic phenomenon, in particular not dependent on details of how the residual resistivity is extracted and is qualitatively robust against details of the parallel resistor model used to extract the resistance of the topmost layer.

This scaling between $T_c$ and $\rho_0$ as both quantities are tuned via Cs deposition is the main result of this manuscript, and does not appear to be limited to this particular Cs surface doped system. A similar scaling between $T_c$ and $\rho_0$ is also observed in monolayer FeSe / $SrTiO_3$ as a function of progressive annealing (Fig. 3f-g) where $\rho_0$ is initially high and $T_c$ is low, but with subsequent annealing, $\rho_0$ falls to a minimum of 1.2 kΩ and $T_{c0}$ increases to a maximum of 31.4 K. The key distinction between monolayer FeSe / $SrTiO_3$ and surface-doped FeSe is that in the monolayer FeSe / $SrTiO_3$, the doping level largely is fixed throughout the annealing process, as the Luttinger volume, as measured by ARPES, changes minimally with annealing[25] (SI Appendix, Fig. S11), and thus the primary variable is solely the level of disorder. Given that such similar $T_c$ - $\rho_0$ scaling is observed in both systems (Fig. 3g), it follows that the same mechanism may be at play in the surface doped system, and furthermore that the driving force behind the $T_c$ - $\rho_0$ scaling is not primarily a carrier doping effect. The presence of the $SrTiO_3$ interface and additional sources of interfacial disorder may explain the different slopes in the $\rho_0$- $T_c$ relation (Fig. 3g). Nevertheless, it is very clear that there is the same, inverse linear relationship between $\rho_0$ and $T_c$ in both systems. Furthermore, in intercalated and electrostatically gated FeSe[5–7,13], $\rho_0$ and $T_c$ also follow the same qualitative relationship all the way up to the highest doping levels reported (typically up to optimal doping). One notable study on $Li^+$-intercalated FeSe flakes exhibits the same qualitative correlation between $\rho_0$ and $T_c$ even slightly past optimal doping, before the sample becomes highly insulating[10].

It is interesting to note that the relationship between $\rho_0$ and $T_c$ as shown in Fig. 3g is so robust, given that $\rho_0$ is a composite and not fundamental quantity. The residual resistivity is determined by both the elastic scattering rate ($\tau_0$) and details of the electronic structure via the carrier density ($n$) and effective mass ($m$), ($\rho_0 = \frac{m}{ne^2\tau_0}$). The fact that $\rho_0$ determines $T_c$ for both underdoped and overdoped compositions regardless of the evolution of the electronic structure suggests that $\tau_0$ may be the primary control parameter for superconductivity, and not changes in the electronic structure with doping. This is qualitatively confirmed by converting the data shown in Fig. 3g to scattering times using the measured carrier densities, effective masses, and residual resistivities (SI Appendix, Fig. S10) which show a similar general trend of longer scattering times corresponding to higher critical temperatures with the highest $T_{c0}$ values coinciding exactly with the longest scattering times. This is also consistent with the quasiparticle widths extracted from the ARPES data (SI Appendix, Fig. S10) which show that the scattering time measured from the spectroscopic linewidths exhibits a non-monotonic doping dependence, with the sharpest spectra and longest scattering times near optimal doping.

**Discussion**

In most systems exhibiting a superconducting dome, the parent compound is an insulator or bad metal, where doped carriers increase the carrier density and screening, which can offset the disorder induced by impurity dopants. Figures 4a-c show the domes of the Fe-pnictide[36-38], cuprate[39,40], and oxide 2DEG[41–43] superconductors, plotted together with the evolution of the normal state resistance, $R_N$, with doping, where $R_N$ is taken just above $T_c$ (here we use $R_N$ instead of $\rho_0$ due to the availability of data in the existing literature). In all of these cases, $R_N$ decreases monotonically across the entire phase diagram and $T_c$ appears largely insensitive to $R_N$. This is in sharp contrast to the highly non-monotonic behavior of $R_N$ (or equivalently $\rho_0$) and its correlation with $T_c$ that we observe in FeSe (Fig. 4d). We emphasize again that along the doping axis, many factors that could plausibly affect both $\rho_0$ and $T_c$ are changing, most obviously the elastic scattering rate and the carrier density. With regards to the non-monotonic evolution of $\rho_0$, one possibility is a natural minimum which occurs as a result of the counterbalancing effects of increasing dopant disorder and increasing screening of impurity scattering as a function of increased doping. However, this picture cannot explain the fact that the minimum of $\rho_0$ coincides exactly with the highest $T_c$ optimal doping, nor the precise quantitative relationship between $\rho_0$ and $T_c$ for doping levels on opposite sides of the phase diagram with very different carrier densities. Furthermore, as discussed previously, the same $T_c$ - $\rho_0$ scaling also occurs in the monolayer FeSe / $SrTiO_3$ system as a function of annealing which rules out changes in doping as the primary factor. We emphasize that although the non-monotonic evolution of $\rho_0$ with Cs deposition allows for us to conclusively and causally link $T_c$ and $\rho_0$ via the relationship shown in Fig. 3g, such a non-monotonic dependence is not required for the relationship to still hold, as seen in the case of monolayer FeSe / $SrTiO_3$.

In other systems, the superconducting dome often appears near a quantum critical point of a magnetic or other competing phase whose fluctuations are believed to be involved in the superconducting pairing mechanism. The combination of the ordered phase competing with superconductivity and the fluctuations of the phase contributing to the superconducting pairing naturally leads to the presence of a superconducting dome. In electron-doped FeSe, a completely unique mechanism may be at play, where the superconductivity is controlled by impurity scattering from disorder, with carrier doping not playing a direct role in determining $T_c$, and thus the superconducting phase diagram may be of a qualitatively different nature than that of other unconventional superconductors. We speculate that in electron-doped FeSe, the interplay between the two competing doping-controlled factors of increased screening versus increased disorder leads to a non-monotonic evolution of its elastic scattering rate, generating the superconducting dome without phase competition or the presence of a quantum critical point. Future experiments combining electron doping with other tuning parameters such as high-energy electron irradiation or high pressure and tracking how the superconducting dome and inelastic/elastic scattering rates evolve across the entire phase space will be useful for probing the correlation between changes in the inelastic scattering rate and superconductivity traditionally associated with quantum critical phenomena, as well as providing a greater range of parameters over which to correlate superconductivity and the residual resistivity.

These results clearly demonstrate the primary role of the residual resistivity in determining $T_c$ in electron-doped FeSe, regardless of the doping level. Nevertheless, it is well established that increasing the carrier density is still essential for transmuting bulk FeSe with a low $T_c$ and a multiband Fermi surface, with both hole and electron pockets, into a high-$T_c$ phase with only electron pockets. We emphasize that our result does not contradict this fact that electron doping is required to induce a high $T_c$ in FeSe. Rather, our results indicate a possible scenario (Fig. 4e) where once the high-$T_c$ phase is reached, it becomes largely insensitive to further changes in carrier doping, and the superconducting dome is generated as the doping parameter traces a nonlinear path through the phase space of impurity scattering and carrier density (black path in Fig. 4e).

One open question remains as to why $T_c$ exhibits such a strong sensitivity to disorder. In superconductors with a sign-changing order parameter, $T_c$ can be heavily suppressed by impurity scattering, e.g., $Sr_2RuO_4$[44]. The low $T_c$ bulk phase of FeSe is also suppressed by disorder[45,46], likely due to the effect of interband impurity scattering on the $s^{+-}$ order parameter. The results presented here suggest a sign-changing superconducting order parameter in electron-doped FeSe as well. Although ARPES measurements on monolayer FeSe / $SrTiO_3$ do not observe nodes in the superconducting gap[47], other possibilities for the electron-doped FeSe superconducting order parameter include an $s^{+/-}$ structure where the sign change occurs between the inner and outer pockets formed by hybridization between the two nearly degenerate electron pockets at the $M$ point (Fig. 4e). While there is consensus that the superconducting order parameter of bulk FeSe is of $s^{+/-}$ symmetry, similar to most Fe-based superconductors, the sign change of the order parameter in those materials is between the hole pockets at the $\Gamma$ point and electron pockets at the $M$ point. In electron-doped FeSe, including monolayer FeSe / $SrTiO_3$, only the electron pockets at $M$ cross the Fermi level. The proposed $s^{+/-}$ character of the superconducting order parameter in the electron-doped, high-$T_c$ system is qualitatively different, changing sign between only electron pockets at $M$. In this case, scattering from disorder even with very low momentum transfer would connect the two pockets with opposite sign and act as pair-breaking scattering events, leading to a very sensitive dependence to disorder as our results suggest. We also note that in both the monolayer FeSe / $SrTiO_3$ and surface doped FeSe, superconductivity emerges close to the pair quantum of resistance ($R_Q$), indicating that reduced dimensionality could also play a role in increasing sensitivity to disorder and 2D phase fluctuations regardless of the pairing symmetry, although this may not be relevant in systems further from the 2D limit such as intercalated systems.

This systematic study of superconductivity across the entire superconducting dome of electron-doped FeSe clearly demonstrates that $T_c$ scales with $\rho_0$ across the phase diagram, suggesting that superconductivity may be fundamentally disorder-limited once the high-$T_c$ electron-doped region of the phase diagram is reached. Given that surface electron doping combined with *in situ* electrical transport and photoemission as employed here provides the most direct and cleanest means for probing superconductivity and tuning doping in a uniquely controllable and continuous fashion, this behavior appears not to be isolated to Cs-surface doped thin films, but generic to all electron-doped FeSe, in which case the path to generating a higher $T_c$ is possibly reduced to increasing the cleanliness of the electron-doped material.

## Materials and Methods

### Thin film synthesis

Epitaxial thin films of FeSe were grown on undoped $SrTiO_3$ (001) (10 mm x 10 mm) substrates using molecular beam epitaxy at a substrate growth temperature of 420 °C and background pressure of $6 \times 10^{-10}$ torr in an adsorption-controlled regime where the nominal flux ratio Se:Fe was 5:1 and were monitored during growth using reflective high energy electron diffraction (RHEED). Prior to film growth, the $SrTiO_3$ substrates were annealed at 600 °C for approximately 6 hours. Under these conditions, the growth rate is approximately 36 seconds per FeSe monolayer. Samples were then transferred through ultrahigh vacuum for *in situ* ARPES and electrical transport measurements.

### Alkali surface doping

A key for this experiment was the precise control over the surface doping, which was achieved by evaporating Cs from a Cs-In alloy source[31] in an effusion cell, which provides a highly stable and reproducible atomic flux of Cs that can be independently measured and calibrated using a QCM, in contrast to conventional strip-style alkali metal dispensers. The QCM calibration was performed at liquid nitrogen temperatures in a vacuum better than $5 \times 10^{-10}$ to maximize the adsorption of Cs. To facilitate uniform random doping, as opposed to phase separation into ordered overlayers, the Cs was deposited when the sample temperature was $T = 8$ K. The Cs dose was controlled by opening a shutter for a typical dose time between 5 – 60 seconds at a typical deposition rate of 0.0012 e/Fe/sec.

### Electrical transport

*In situ* resistivity measurements were performed using a custom-built UHV four-point transport probe with a base temperature of 7 K and a base pressure of $7 \times 10^{-11}$ torr. Contact is applied directly to the film by depressing Au-plated spring-loaded probes onto Au contacts evaporated onto the film after synthesis using a shadow mask in a 4-point van der Pauw geometry, with a nominal instrumental contact spacing of 7 mm. Resistance measurements are taken using a Keithley 6221 / 2182A current source–voltmeter combination in delta mode with a typical applied current of 20 − 30 μA. Measurements were performed in both warming and cooling cycles to verify that the Cs concentration at the surface did not change upon warming. It was found that the resistivity measurements were completely repeatable when keeping the sample temperature below 80 K, whereas hysteresis was observed above 80 K.

**Extraction of the surface doped layer resistance**

The layered and highly anisotropic nature of FeSe suggests that modeling the film as a stack of individual FeSe layers as resistors adding in parallel is a reasonable approximation. Assuming such a model for a 10 u.c. thick film and given the measured sheet resistance of the undoped stack ($R_{10\ layer}^{undoped}$) and the measured sheet resistance of the doped stack ($R_{10\ layer}^{doped}$), the extracted sheet resistance of the doped surface layer ($R_{1\ layer}^{doped}$) is given by:

$$R_{1\ layer}^{doped} = \left( \frac{1}{R_{10\ layer}^{doped}} - \frac{9}{10} \frac{1}{R_{10\ layer}^{undoped}} \right)^{-1}$$

**ARPES**

ARPES measurements were taken with a VG Scienta R4000 electron analyzer equipped with a VUV5000 helium discharge lamp using He-I photons at 21.2 eV. The base pressure in the ARPES system is $5 \times 10^{-11}$ Torr. The energy resolution is nominally set at 12 meV. To avoid sample charging during ARPES measurements, the film is grounded using a retractable contact pin built onto the sample manipulator. Surface doping during ARPES was achieved via flowing current through alkali metal dispensers (SAES) close to the sample surface with the sample held at the base temperature $T = 7$ K. Doping levels were extracted from the ARPES data by calculating the Luttinger volume using the Fermi momentum ($k_F$) values of the electron pockets at the $M$ point and assuming circular pockets. The optimal doping levels extracted from the Luttinger volumes and from QCM measurements showed good agreement.

**Author Contributions:** P.T.M., B.D.F, and K.M.S. conceived the research and designed the experiment. P.T.M, C.J.M, Y.T., and B.D.F. synthesized the FeSe films and performed the surface doping, in situ resistivity, and ARPES measurements. The data were analyzed and interpreted by P.T.M., B.D.F., and K.M.S. D.G.S. developed the Cs-In source for the surface doping. P.T.M. and K.M.S. wrote the manuscript, with contribution from all authors.

**Competing Interest Statement:** Authors declare they have no competing interests.

**Classification:** Physical Sciences

**Keywords:** Superconductivity, Iron-based superconductors, ARPES

**Acknowledgments**

This work was primarily supported by the Air Force Office of Scientific Research through Grants No. FA9550-21-1-0168 and FA9550-23-1-0161. Additional support was provided through the National Science Foundation through DMR-2104427 and the Platform for the Accelerated Realization, Analysis, and Discovery of Interface Materials (PARADIM) under Cooperative Agreement No. DMR-2039380 and the Gordon and Betty Moore Foundation's EPiQS Initiative through Grant Nos. GBMF3850 and GBMF9073. P.T.M. would like to acknowledge support from the Klarman Postdoctoral Fellowship.

## Figures and Tables

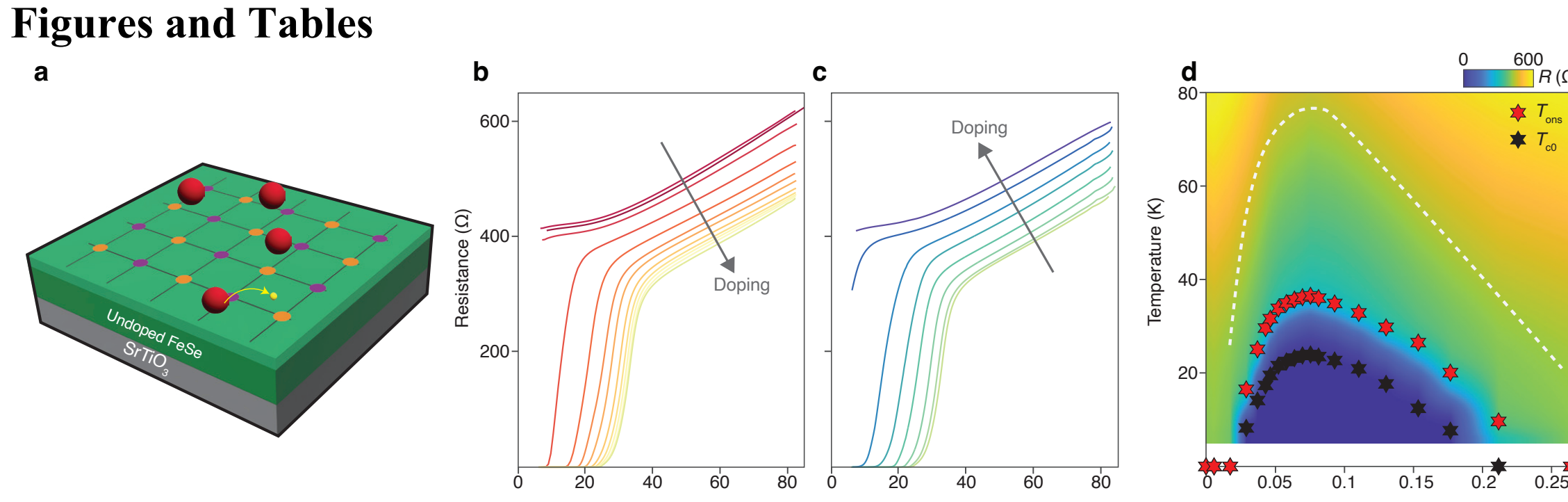

**Figure 1.** Electronic transport as a function of doping via Cs surface deposition. (a) Schematic of 10 u.c. thick FeSe film grown on $SrTiO_3$. Deposited Cs atoms (red) dope electrons into the surface layer (gold) of the FeSe film. Temperature dependence of the sheet resistance as a function of Cs deposition for underdoped (b) and overdoped (c) doping levels. (d) Temperature-doping color map of the sheet resistance and doping dependence of the superconducting transition temperatures $T_{ons}$ and $T_{c0}$. White dashed line is a guide to the eye highlighting the low resistance region of the normal state resistance above the superconducting dome.

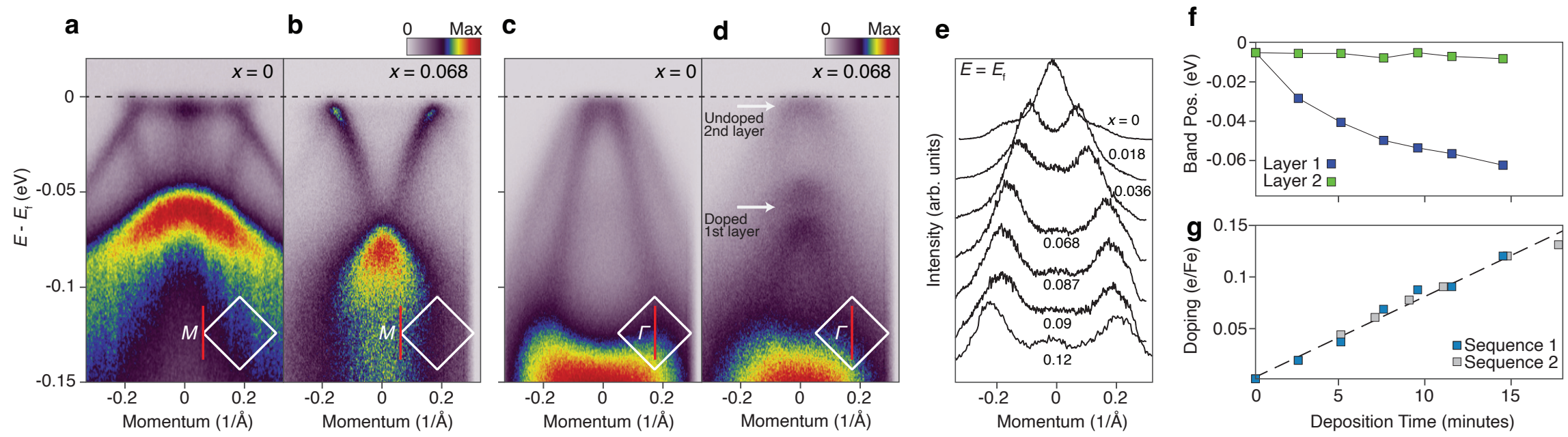

**Figure 2.** Evolution and surface confinement of doping observed in ARPES. (a - d) ARPES spectra at $T = 7$ K at the $M$ point for the as-grown sample (a) and a near optimally doped sample (b), and at Γ for the as-grown undoped sample (c) and the near optimally doped sample (d). The white arrows in (d) highlight the hole pocket for the undoped second layer and doped surface layer. (e) Momentum distribution curves taken at $k_F$ at the $M$ point as a function of increased doping. (f) Band maxima of the hole pockets for the first and second surface layers. (g) Electron doping extracted from the Luttinger volumes as a function of Cs deposition time for two separate doping series. The dashed line is a linear fit to the combined datasets.

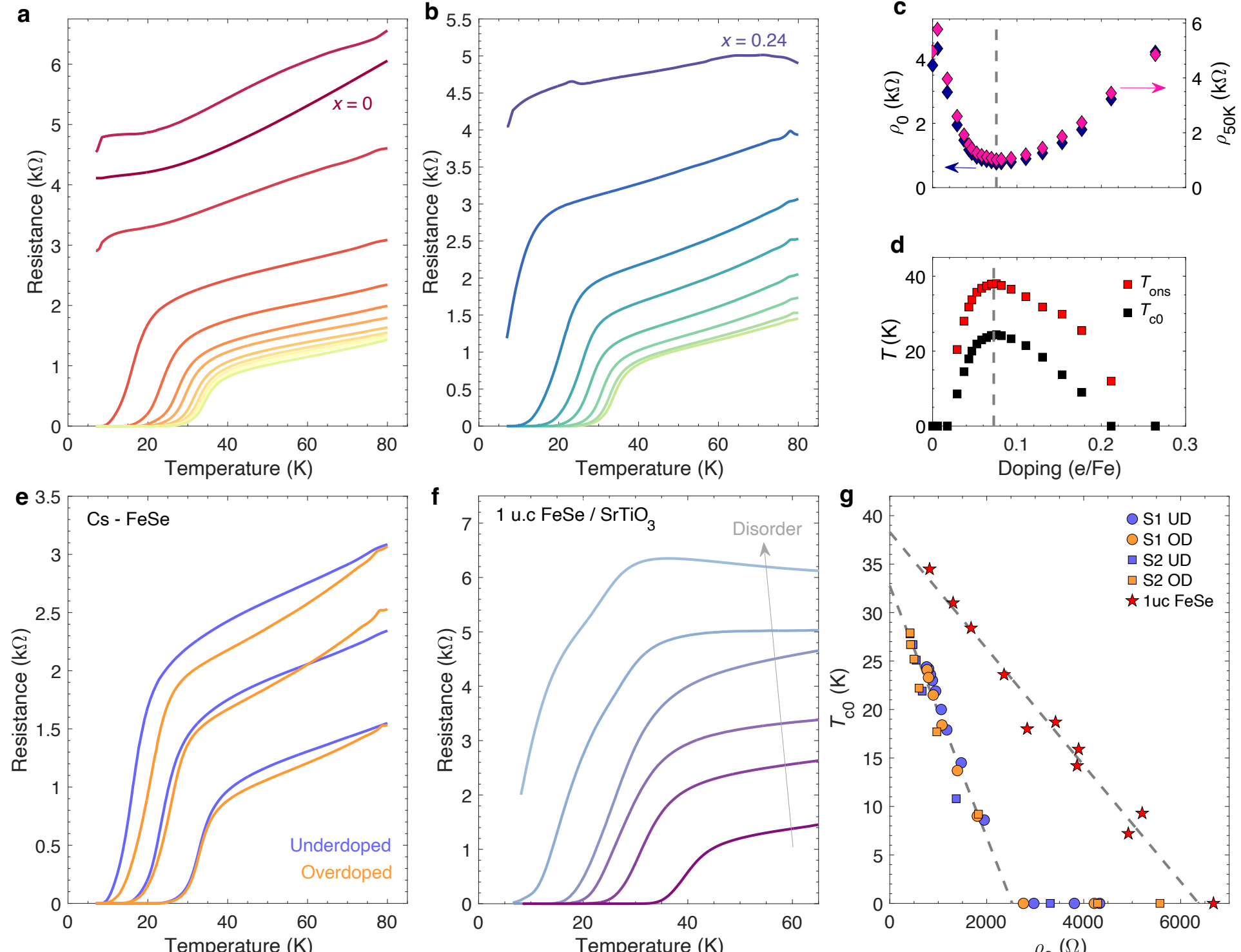

**Figure 3.** Extracted sheet resistance of the doped surface layer. (a, b) Temperature dependence of the extracted resistivity of the doped surface layer for underdoped (a) and overdoped (b) doping levels. (c) Extrapolated residual resistivity $\rho_0$ (left axis) and resistivity at $T = 50$ K (right axis) as a function of electron doping. **(**d) Superconducting critical temperatures $T_{ons}$ and $T_{c0}$ as a function of electron doping. The dashed line in (c, d) marks the position of optimal doping $x_{opt}$. (e) Comparison of resistivity curves for pairs of under- and overdoped samples with similar values of $T_{c0}$. (f) Resistivity as a function of temperature for monolayer FeSe / $SrTiO_3$ samples with varying residual resistivities. (g) $T_{c0}$ as a function of $\rho_0$ for both the Cs deposited (two different doping series) and monolayer FeSe systems. Dashed lines are linear fits to each respective dataset.

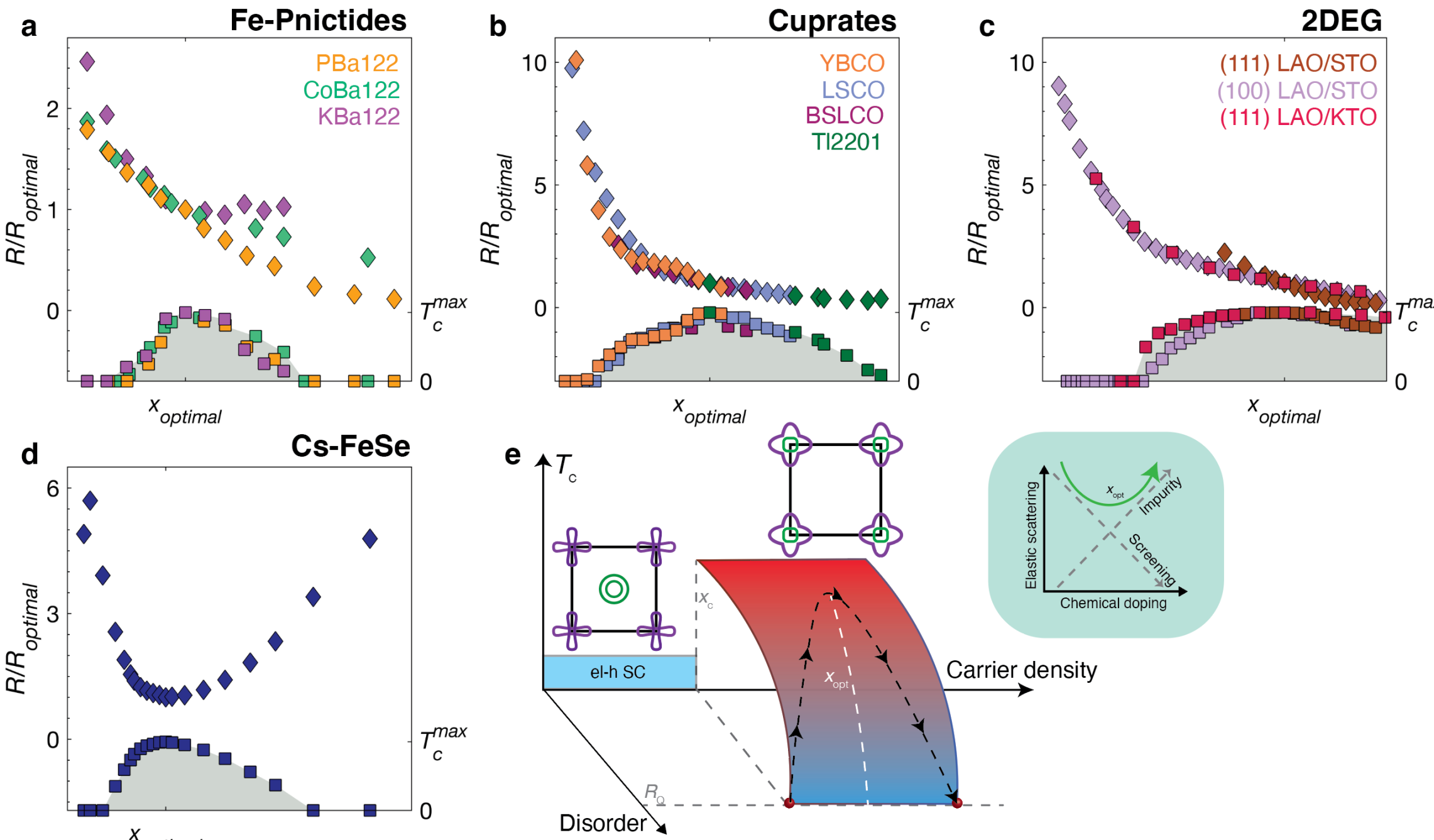

**Figure 4.** Comparison of systems exhibiting superconducting domes. Normal state resistivity (top left axis, diamond markers) and superconducting transition temperature (bottom right axis, square markers) as a function of tuning parameter for (a) Fe-pnictide, (b) cuprate, (c) gated 2D electron gas, and (d) Cs surface doped FeSe superconductors. The resistance data has been normalized by the value of optimal doping for each compound and the horizontal axis has been scaled such that the superconducting domes for different compounds within a family overlap. (e) Schematic phase diagram for the $T_c$ of electron-doped FeSe as a function of disorder and carrier density. Green and purple colors on the Fermi surface sheets indicate the sign of the superconducting order parameter.